\documentclass[aps,prl,reprint]{revtex4-2}
\usepackage{comment}

\usepackage{amsmath,amssymb,hyperref,mathrsfs}
\usepackage{braket,bm,bbm}
\PassOptionsToPackage{normalem}{ulem}

\usepackage[makeroom]{cancel}
\usepackage{graphicx}

\usepackage[usenames,dvipsnames]{xcolor}
\hypersetup{colorlinks=true, citecolor=Plum, linkcolor=Plum,
urlcolor=Plum}


\usepackage{subfigure}
\begin{document}
    
\title{Radiating black holes in general relativity need not be singular}
\author{Francesco Di Filippo}
\affiliation{Institut f\"ur Theoretische Physik, Max-von-Laue-Str.1, 60438 Frankfurt, Germany\\ $\,$}

\begin{abstract}
    It is common knowledge that black holes necessarily contain a region where general relativity breaks down, due to the inevitable formation of either a curvature singularity or a Cauchy horizon. In this work we challenge this view by analyzing a charged spherically symmetric black hole formed through gravitational collapse and evaporating via Hawking radiation. We show that the electromagnetic repulsion and the violation of energy conditions due to the presence of Hawking radiation can be sufficient to avoid the formation of both a singularity and a Cauchy horizon. We argue that a similar mechanism may apply to astrophysical black holes in which the role of the electric charge is replaced by the angular momentum.
\end{abstract}

\maketitle
\section{Introduction}
\noindent Black holes are some of the most fascinating objects in the universe as they constitute both one of the greatest successes of general relativity and its biggest limitation. While tests of the exterior geometry are in perfect agreement with the prediction of the theory, a series of fundamental results, starting from the famous Penrose's singularity theorem \cite{Penrose:1964wq}, proves that black hole interiors cannot be fully described by the theory. Under the assumption of the null convergence conditions (which in general relativity coincide with the null energy conditions) \cite{Poisson:2009pwt}, the theorem tells us that any spacetime with a trapped region must be geodesically incomplete \cite{Penrose:1964wq}. This can happen because of the formation of either a curvature singularity or of a Cauchy horizon. 
Subsequent theorems~\cite{Hawking:1970zqf,Hawking:1973uf,Senovilla:1998oua}  extended the analysis by modifying some assumptions and obtained similar results. 

Attempts to resolve this theoretical conundrum motivated a field of research that has gained a lot of popularity in recent years (see \cite{Carballo-Rubio:2025fnc} for a recent review) and several possibilities have been put forward. It is widely accepted that the resolution of this consistency problem requires either the modification of the gravitational interaction at high energy, possibly due to quantum gravity effects or the introduction of exotic forms of matter. 

In this work, we challenge this perspective. While a violation of the null energy condition is necessary, we note that we do have a source of such a violation which is far from being exotic physics. In fact, black holes evaporate via the emission of Hawking radiation \cite{Hawking:1974rv,Hawking:1975vcx}, a process which necessarily violates the null energy conditions. 

We show that a simple model of gravitational collapse within general relativity can lead to a perfectly regular spacetime without Cauchy horizons if the role of Hawking evaporation is properly taken into account. 
In particular, we show that the gravitational collapse of charged matter can lead to a black hole without a central singularity thanks to both the electrostatic repulsion and the violation of energy conditions provided by Hawking radiation.

\section{Gravitational collapse}
\noindent In this section we review some well-known facts about the gravitational collapse in general relativity. In particular, we are interested in the difference between a neutral and a charged matter distribution. For the time being, we do not include the effect of Hawking radiation, that is, the black hole is eternal once formed. 

\begin{figure}
    \centering
    \includegraphics[width=0.36\linewidth]{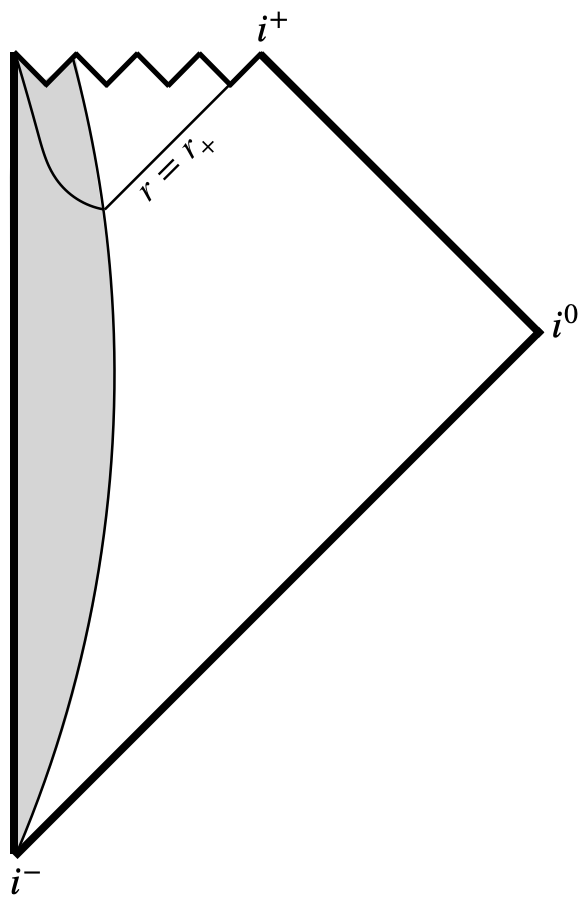}
    \caption{Gravitational collapse into a Schwarzschild black hole.}
    \label{Fig:Opph_Snei}
\end{figure}

We start by considering the collapse of uncharged spherically symmetric matter into a Schwarzschild black hole (the simplest case is that of pressureless uniform dust~\cite{PhysRev.56.455}). 
The causal structure of this spacetime is depicted in Fig.~\ref{Fig:Opph_Snei}. Because of the uniqueness theorems \cite{PhysRev.164.1776}, the exterior geometry is given by the Schwarzschild spacetime: 
\begin{equation}
    ds^2=-\left(1-\frac{2M}{r}\right)dt^2+\left(1-\frac{2M}{r}\right)^{-1}dr^2+r^2d\Omega^2\,,
\end{equation}
which has a single horizon at $r=2M$. The details of the interior geometry depend on the specific matter under consideration. However, there are some elements that are universal. Most importantly, an inner horizon forms inside the matter distribution. As the outer geometry can only have a single horizon, the inner horizon must remain within the matter distribution. In turn, this implies that the collapsing matter cannot escape the trapping region and the collapse cannot be stopped. Eventually the radii of both the inner horizon and the collapsing matter distribution shrink to zero. When this happens, we know that a curvature singularity has to form (see e.g.~\cite{Carballo-Rubio:2019fnb}). Such singularity must be spacelike as it has formed inside a trapped region. 

\begin{figure}
    \centering
    \subfigure[$\,$]{\includegraphics[width=0.33\linewidth]{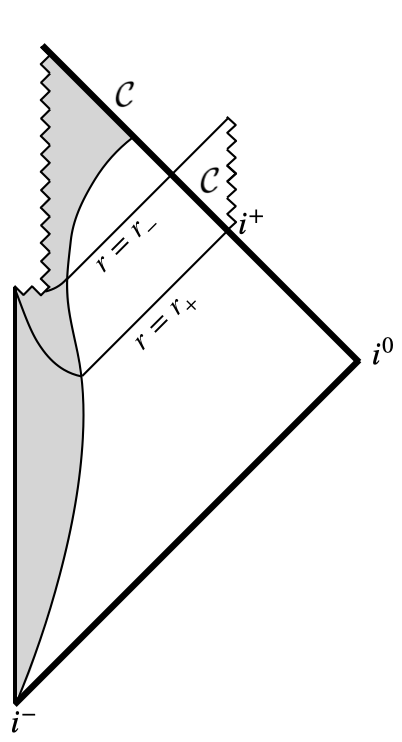}\label{subfig:RN_collapse_singular.png}}
    \hspace{2cm}
   \subfigure[$\,$]{ \includegraphics[width=0.3\linewidth]{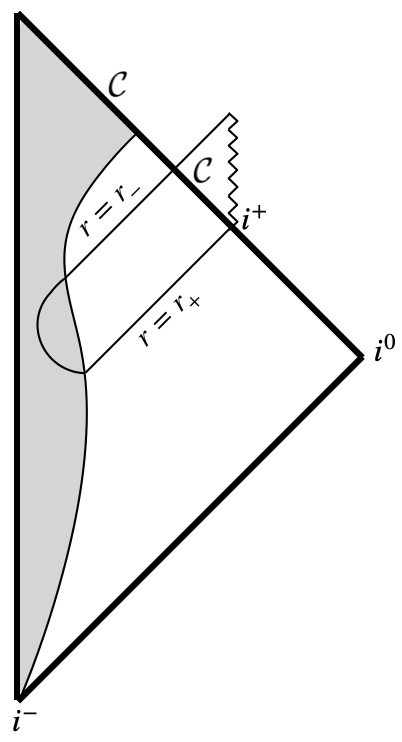}\label{subfig:RN_eternal_reg}}
    \caption{Gravitational collapse leading to the formation of a Reissner--Nordstr\"om black hole. In the left diagram, the inner horizon collapses and reaches $r=0$. At this point a spacelike singularity is formed. Once the matter shrinks inside the inner horizon for the exterior geometry, an untrapped region is formed and eventually the singularity becomes timelike. Part of the matter forming the black hole collapses into the singularity while the rest can bounce and cross the Cauchy horizon.  In the right diagram, the radius of the collapsing matter shrinks within the inner horizon radius of the exterior geometry. All the matter can bounce inside the trapping horizon without forming a singularity inside the matter distribution. Eventually, the matter crosses a Cauchy horizon. }
    \label{fig:RN_eternal}
\end{figure}

Let us now describe what changes if we consider a charged matter distribution.
Also this case is quite well studied. The exterior metric is once again fixed by uniqueness results \cite{1968CMaP} and is given by the eternal Reissner--Nordstr\"om spacetime
\begin{equation}
\begin{array}{rl}
        ds^2=&-\left(1-\frac{2M}{r}+\frac{Q^2}{r^2}\right)dt^2+\left(1-\frac{2M}{r}+\frac{Q^2}{r^2}\right)^{-1}dr^2+\\
        \\
        &+ r^2d\Omega^2\,.
        \end{array}
\end{equation}
This spacetime has both an inner and an outer horizon, located at
\begin{equation}
    r_\pm=M\pm\sqrt{M^2-Q^2}\,.
\end{equation}
It is instructive to look at the causal structure of this spacetime. 
Both horizons are null and last for infinite time (as seen from an outside observer). Asymptotically, outside the matter distribution a Cauchy horizon $\mathcal{C}$ forms. Let us now turn our attention to the geometry inside the matter distribution.
The main difference is due to the presence of the inner horizon in the outer geometry. The matter distribution undergoing the collapse eventually has to cross the inner horizon of the exterior geometry leaving a non-trapped region in the interior where the matter distribution can undergo a bounce.
This can happen in two possible ways both illustrated in Fig.~\ref{fig:RN_eternal}. 
As depicted in Fig.~\ref{subfig:RN_collapse_singular.png}, the inner horizon can shrink faster than the radius of the matter distribution and reach $r=0$ within the matter implying that a spacelike singularity has to form \cite{Carballo-Rubio:2019fnb}.
Once the radius of the collapsing matter gets below the inner horizon for the exterior Reissner--Nordstr\"om spacetime $R<r_-$, an untrapped region is formed and eventually the singularity becomes timelike. Part of the matter collapses into the singularity, some matter can bounce and cross the Cauchy horizon.
The second possibility is shown in Fig.~\ref{subfig:RN_eternal_reg}. In this case, the radius of the collapsing matter shrinks fast enough with respect to the segment of the horizon in the interior geometry so that the latter does not reach $r=0$ and it merges with the inner horizon of the exterior Reissner--Nordstr\"om geometry. In this case, the matter can bounce and cross the Cauchy horizon without the need of forming the singularity inside the matter distribution.  

Which of these two scenarios happens in practice  depends on the details of the matter undergoing collapse (see chapter 10 of \cite{Hawking:1973uf}). For instance, the first scenario is bound to happen if enough uncharged matter is present. On the other hand, the second scenario arises for some simple matter distributions. For example, this is the case both for thin shells \cite{Cruz1967GravitationalB} and for  pressureless uniform matter distributions (with arbitrary low charge)~\cite{Bekenstein:1971}. In particular, already in the early 1970s, Bekenstein writes ``\textit{after each shell
of matter has crossed its inner horizon, it ``bounces" and reexpands into a region of spacetime (Carter-Penrose diagram) different from the one in which the collapse originated. Thus the
central singularity is avoided.}"~\cite{Bekenstein:1971}.

In this letter, we study the possibility of producing black holes without any singularity. Therefore, in what follows we focus on the second possibility. 

Before discussing the role of evaporation, let us comment on how the charged and uncharged gravitational collapses fulfill the consequences of Penrose's singularity theorem~\cite{Penrose:1964wq}. Both cases form a trapped surface and both clearly respect the null energy conditions for physically realistic matter distributions. Penrose's singularity theorem, therefore, predicts that the spacetime is geodesically incomplete. For the non-charged case, incomplete geodesics end at the singularity. For the charged case, geodesics are incomplete as they reach the Cauchy horizon for finite values of the affine parameter. We emphasize that the bounce of the collapsing matter inside the inner horizon and the avoidance of the singularity inside the matter distribution (left side of the Penrose diagram in Fig.~\ref{subfig:RN_eternal_reg}) is not in contradiction with the conclusions of the singularity theorems. Indeed, this is achieved within classical general relativity without violations of energy conditions.

\section{The role of evaporation}
\noindent Let us now discuss how Hawking evaporation modifies the scenarios illustrated above. The most important feature is that Hawking radiation violates the null energy conditions. This allows, among other things, the outer horizon to be timelike \cite{Hayward:1993mw,Ashtekar:2004cn}. 

For the non-charged case, the analysis is quite straightforward. A Schwarzschild black hole evaporates with temperature $T_S$ given by \cite{Hawking:1974rv,Hawking:1975vcx} (in units $c=G=k_B=1$)
\begin{equation}\label{eq:Schw_temp}
    T_S=\frac{\hbar}{8\pi M}\,.
\end{equation}
As the mass of the black hole shrinks, due to the backreaction of the evaporation, the temperature increases and the black hole evaporates completely in finite time. Whether it leaves a singular or regular remnant depends on the physics at play at the end of the evaporation, where the temperature \eqref{eq:Schw_temp} becomes Planckian. Regardless, a singularity definitely forms inside the trapped region. 

A very different scenario can arise in the charged case. As previously mentioned, we restrict our attention to the physical scenario in which no singularity forms in the interior (e.g. pressureless dust or thin shells)

The charged black hole radiates with temperature \cite{Fabbri:2000es,Sorkin:2001hf,Diba:2002hb}
\begin{equation}\label{eq:RN_temp}
    T_{RN}=\frac{\hbar}{4\pi}\frac{r_+-r_-}{r_+^2}\,.
\end{equation}
In this case, the temperature vanishes as we approach the extremal case. What happens at the end point of evaporation depends on how we model the final stages of evaporation. In particular, a radiating charged black hole loses both energy and electric charge. Therefore, we can model the evaporation simply by promoting the mass and the electric charge to time dependent functions and at each time the evaporation proceeds according to Eq.\eqref{eq:RN_temp} with time dependent values for the inner and outer horizon. Generically, the extremal case would be reached in infinite time (see e.g. \cite{Fabbri:2000es,Sorkin:2001hf,Diba:2002hb,Carballo-Rubio:2018pmi}). However, this does not take into account that the distance between the horizons becomes sub-Planckian in finite time and it is unlikely to expect that no other effect comes into play to determine the dynamics of the inner and outer horizon. For instance, the mass inflation instability will likely play a very relevant role \cite{Barcelo:2020mjw,Barcelo:2022gii} (we will comment on this point in more details in the next section). 

Therefore, it is instructive to consider all the different possibilities for the end of evaporation and examine the causal structure and the properties of each case. 

\begin{widetext}

\begin{figure}[!htb]
    \centering
    \subfigure[$\,$]
    {\includegraphics[width=0.14\linewidth]{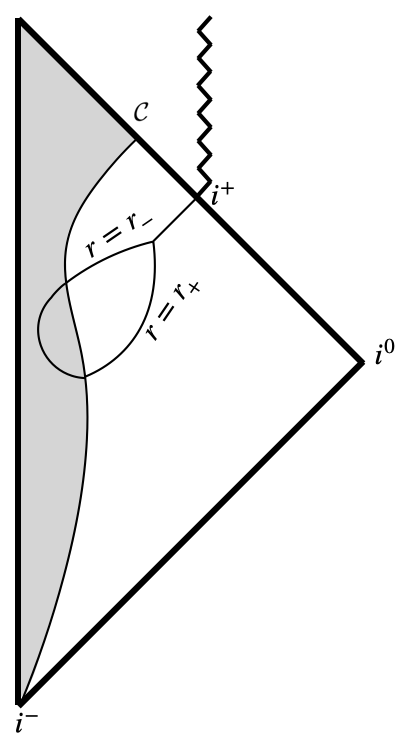}\label{subfig:RN_evaporation_remnant}}
            \hspace{1cm}
  \subfigure[$\,$]{\includegraphics[width=0.14\linewidth]{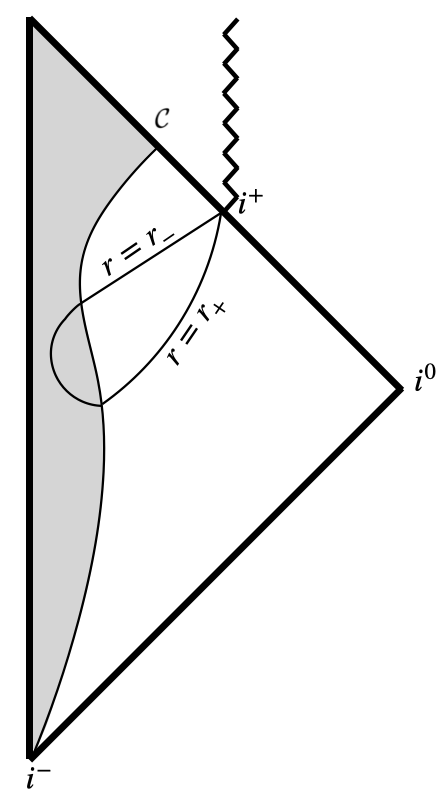}\label{subfig:RN_evaporation_scri}}
    \hspace{1cm}
    \subfigure[$\,$]{\includegraphics[width=0.14\linewidth]{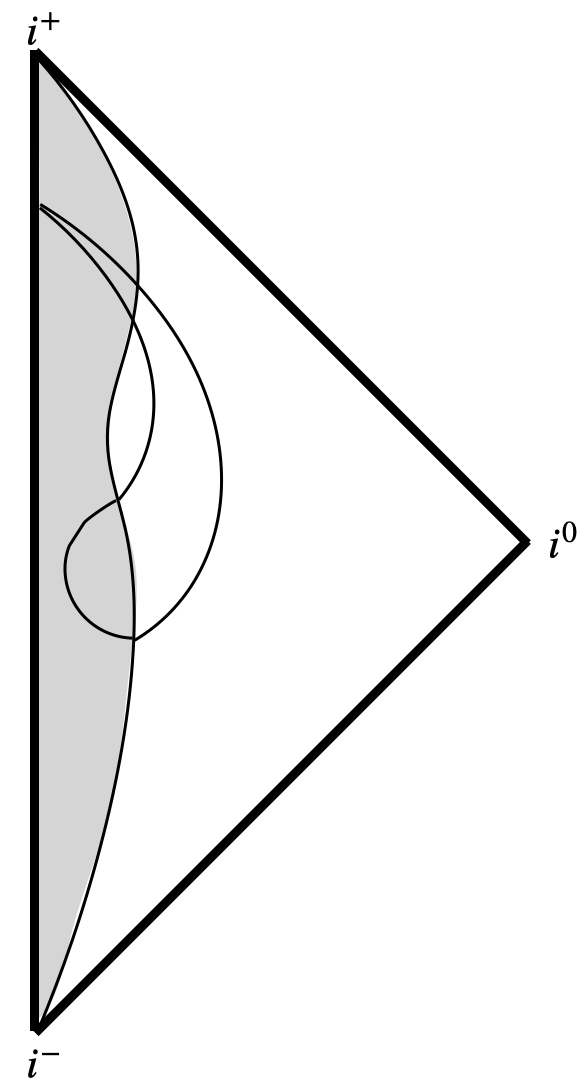}\label{subfig:RN_evaporation_0}}
    \hspace{1cm}
    \subfigure[$\,$]
    {\includegraphics[width=0.14\linewidth]{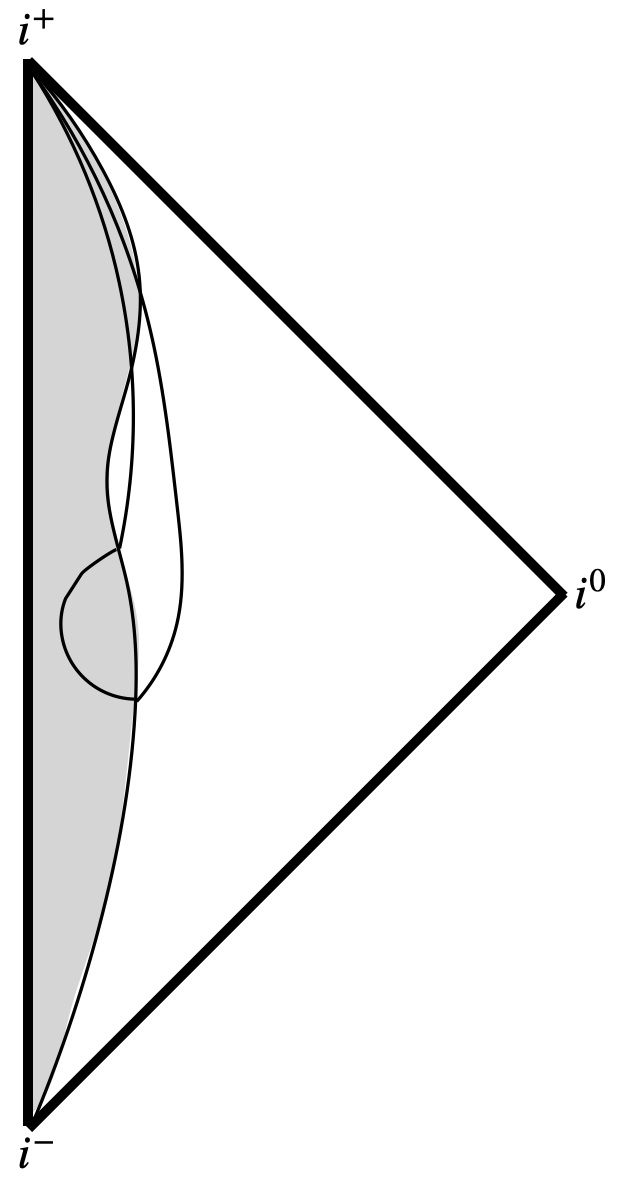}\label{subfig:RN_evaporation_i0}}
\hspace{1cm}
  \subfigure[$\,$]{            \includegraphics[width=0.135\linewidth]{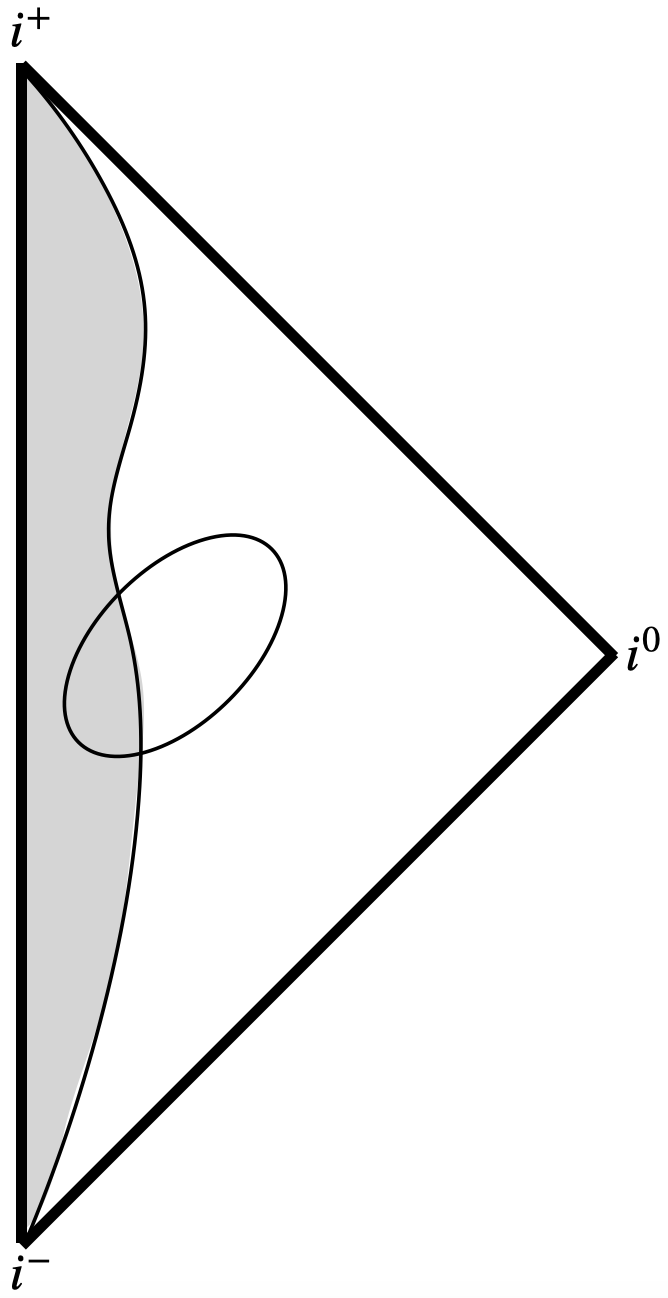}\label{subfig:RN_evaporation_R}}
    \caption{Different possible scenarios for the end-point of the gravitational collapse. (a) Formation of an extremal remnant in finite time; (b) asymptotic formation of an extremal black hole; (c) complete evaporation in finite time; (d) complete evaporation in infinite time; (e) formation of a horizonless remnant. Out of the five scenarios the first two lead to the formation of a Cauchy horizon and hence breakdown of predictability, while the other cases lead to a regular spacetime with no Cauchy horizons. }
    \label{fig:RN_Evaporating}
\end{figure}
\end{widetext}

\paragraph{Case a: Extremal horizon remnant.}
The first possibility we consider is the one where an extremal horizon is formed in finite time. The evaporation would stop as the resulting spacetime would have vanishing surface gravity and hence zero temperature. Note, however, that this scenario would violate the third law of black hole thermodynamics~\cite{Bardeen:1973gs}. In this case, the late time geometry is the extremal Reissner--Nordstr\"om geometry. As shown in Fig.~\ref{subfig:RN_evaporation_remnant}, this implies that the spacetime has a Cauchy horizon. Behind the Cauchy horizon, a singularity is also present.  

\paragraph{Case b: Asymptotic extremal horizon.}
In this case, the two horizons merge asymptotically at a non-vanishing value of the radius. We need to determine if the merge at future null infinity or future timelike infinity. 
This can be done by noting that the spacetime approach a static extremal regime for which
\begin{equation}
    r_\pm=M=\left|Q\right|\,
\end{equation}
is a null hypersurface. Therefore, the horizons must merge at asymptotic null infinity. This situation is depicted in Fig.~\ref{subfig:RN_evaporation_scri}. 
As before, a Cauchy horizon and a singularity are present. This type of spacetime would  arise if we ignore the backreaction due to mass inflation at the inner horizon and consider Hawking evaporation via a non-charged field. In this case, the mass decreases to be consistent with the emission of energy and the charge stays constant. The extremal configuration is only reached asymptotically~\cite{Carballo-Rubio:2018pmi}. 

\paragraph{Case c: Complete evaporation.}
Similarly to the first case, this scenario corresponds to a black hole that evaporates completely in finite time. In this scenario, however, the object evaporates completely leaving no remnant. The inner and outer horizon merge at $r=0$. The matter collapsing escapes the trapped region without producing a singularity nor a Cauchy horizon.\footnote{This assumes that the outer and inner horizon reach $r=0$ simultaneously. If the charge is completely radiated away while the mass is still finite, the inner horizon vanishes before the outer one. In this case a singularity would form \cite{Carballo-Rubio:2019fnb}.} This configuration is depicted in Fig.~\ref{subfig:RN_evaporation_0}. 

\paragraph{Case d: Asymptotic evaporation.}
In this scenario, the black hole  fully evaporates asymptotically. A crucial question is whether the inner and outer horizons merge on future null infinity or on future timelike infinity. We can understand this point in the following way. Inside the inner horizon, the hypersurface $r=0$ is clearly timelike. Outside the inner horizon, the spacetime is approximately flat. Therefore any constant radius hypersurfaces $r=\epsilon$ is  timelike for arbitrarily small values of $\epsilon$ (at sufficiently late times to ensure that the hypersurface is outside the outer horizon). By continuity, this implies that the horizons merge on asymptotic timelike infinity $i^+$, as shown in Fig.~\ref{subfig:RN_evaporation_i0}.
The matter distribution can escape the trapped region\footnote{Matter can escape an eternal trapped region as the horizon is timelike due to the violation of the energy conditions.} leaving a regular spacetime without any Cauchy horizon.  An example of this spacetime would arise if we ignore the backreaction due to mass inflation at the inner horizon and consider Hawking evaporation via a charged field. As for the \textit{case b}, the mass decreases to be consistent with the emission of energy. However, in this case also the charge decreases and vanishes asymptotically.

\paragraph{Case e: Horizonless remnant. }
The last possibility consists of a black hole evaporating in finite time leaving a horizonless remnant. The spacetime diagram of this scenario is depicted in Fig.~\ref{subfig:RN_evaporation_R}. Because of the disappearance of the trapped region, there is no Cauchy horizon in this spacetime. The collapsing matter, that would cross the Cauchy horizon in the eternal case, now simply escapes towards asymptotic infinity. It is possible to obtain a concrete realization of this type of spacetime by assigning by hand the evolution of the mass and charge parameter. A similar construction has been discussed for regular black holes geometries \cite{Frolov:2017rjz,Carballo-Rubio:2022nuj}.

All of the cases we have discussed fulfill the requirements of Penrose's singularity theorem as the Hawking radiation provides a source of violation of the energy conditions. The first two scenarios also have a Cauchy horizon and a singularity, even if this was not necessarily required by the theorem.

Crucially, the last three scenarios do not have any breakdown of predictability, i.e., no singularity and no Cauchy horizon.
\section{Open questions}
\noindent There are several open questions that still need to be addressed.

First of all, we have not discussed the interaction between the collapsing matter and the ingoing negative energy flux of Hawking radiation. This is clearly an important piece of the puzzle. If such interaction is discarded, from the causal structures of the different scenarios of Fig.~\ref{fig:RN_Evaporating} it appears that all the matter that collapsed to form the trapped region escapes at the end of evaporation when the black hole is supposed to have very little (if not vanishing) mass.  

A second open question regards the mass inflation instability of the inner horizon. In fact, it is well known that the inner horizon is generically unstable \cite{Poisson:1989zz,Poisson:1990eh}. The instability is usually analyzed in stationary spacetimes, where the inner horizon is also a Cauchy horizon. However, recently it was shown that the instability also applies to dynamical inner horizons~\cite{Carballo-Rubio:2024dca}. The backreaction could push the problem beyond the validity of the semiclassical approximation. Therefore, semiclassical gravity is the correct framework in which to pose the questions raised in this letter, but its validity has to be checked a posteriori.

Moreover, we have analyzed different possibilities for the end-point of the evaporation. We might wonder which one happens in practice. This likely depends on the answer to the previous point. If we only consider Hawking radiation, the evaporation would only end asymptotically \cite{Carballo-Rubio:2018pmi}. However, the backreaction of the instability at the inner horizon will likely play a fundamental role. In fact, there are already analyses showing that the backreaction of semiclassical effect on top of a Reissner--Nordstr\"om black hole can produce the disappearance of the trapped region in finite time~\cite{Boyanov:2025otp}. Therefore, while more investigations are needed, the current analyses tend to point to a scenario in which a horizonless spacetime is formed in finite time, compatible with Fig.~\ref{subfig:RN_evaporation_R}. We might wonder what the timescale of the evaporation would be. Unfortunately, we are not at the stage to answer this question. If the evaporation is mostly driven by Hawking radiation until the two horizons are very close to each other, we can expect the usual evaporation timescale of order $M^3$. However, the mass inflation instability at the inner horizon could potentially lead to a much faster process. Crucially, all these questions can be analyzed numerically in general relativity, no input from physics beyond general relativity is needed.

Additionally, we may wonder if there is an interplay between our analysis and the information loss problem \cite{Hayward:2005ny,Mathur:2009hf}. Given that in this picture there is no formation of any singularity, Cauchy nor event horizons, we expect that the information must be preserved and escape to infinity in the late stage of evaporation. Note, however, that this would require the violation of the area limit for the entanglement entropy \cite{RevModPhys.93.035002,Buoninfante:2021ijy}

Finally, and most importantly, we need to study if a similar mechanism of singularity avoidance would apply to astrophysical black holes. Answering this question requires a careful investigation of the rotating case. However, we expect that the answer will be positive because of the many similarities in the causal structure of Reissner--Nordstr\"om and Kerr black hole. In particular, the electric charge is often discussed as a proxy to understand the behavior of black holes with non-vanishing angular momentum. The centrifugal repulsion would play the role of the electrostatic interaction. However, let us mention that the renormalized semiclassical stress energy tensor for a rotating black hole has a much richer structure than the one for charged black hole. Crucially, the expectation value of the renormalized stress-energy
tensors at the Cauchy horizon of Kerr black holes shows a complex angular dependence~\cite{Fernandes:2023vux,McMaken:2024fvq}.  
\section{Conclusions}
\noindent In this work we have shown that radiating black holes in general relativity need not be singular. 

In fact, we have classified all possible spacetimes  formed via gravitational collapse that evaporate via Hawking radiation based on their causual structure. We have seen that some of the possible scenarios are perfectly regular and contain neither a Cauchy horizon nor a curvature singularity.

A crucial point is that no exotic physics was added by hand. There is no need to modify the dynamics of gravitational interaction or to invoke exotic forms of matter. The combination of the repulsive effect due either to the electromagnetic interaction (for the simple case analyzed here) or to the rotation of spacetime (for a more relevant astrophysical scenario) and the violation of energy conditions due to semiclassical gravity can be enough to avoid the breakdown of the validity of the theory. A classical mechanism of repulsion can produce an inner horizon and a bounce of the collapsing matter. This avoids the formation of the singularity inside the matter distribution. This does not contradict any singularity theorem as a singularity and/or a Cauchy horizon still forms in the other asymptotic region. On the other hand, the role of semiclassical gravity (which is  negligible within the collapsing matter) can prevent the formation of the Cauchy horizon, and the singularity beyond it, by violating the energy conditions and allowing the trapping region to disappear.

There are still several unanswered questions. In particular more careful analyses are needed to understand which of the causal structure in Fig.~\ref{fig:RN_Evaporating} is realized. To this end, it is probably needed to go beyond what is considered a ``standard" modelization of Hawking radiation in the sense that the backreaction of the inner horizon must be under control. Crucially, this seems to be more of a calculational than a conceptual problem as all the physical ingredients are known. 

In conclusion, this work provides a proof of principle that the resolution of the singularity problem does not require exotic physics nor a full theory of quantum gravity, in stark contrast with what is widely assumed. Far from being a strong proof of the realization of any specific scenario, this paper suggests that classical general relativity, supplemented by known semiclassical effects, may already contain the seeds of singularity resolution.

\section{Acknowledgments}
\noindent The day before this manuscript was uploaded, the paper~\cite{Gralla:2025gzl}, where a related idea is discussed, appeared on arXiv. 
There are some differences between the two works. The present one focuses on the singularity problem and tangentially touches issues related to the end of evaporation and the information loss problem. On the other hand,~\cite{Gralla:2025gzl} mostly focuses on the end point of the evaporation and the information loss problem, and at the end it explains the implications for the singularity problem. Furthermore, the cases in Fig.~\ref{subfig:RN_evaporation_remnant},~\ref{subfig:RN_evaporation_0} and~\ref{subfig:RN_evaporation_i0} discussed in the present work were not considered in~\cite{Gralla:2025gzl}.  However, the overlap of results is substantial.\\
I am grateful to Luca Buoninfante, Francesco Del Porro and Costantino Pacilio for useful discussions in the early stages of this project. I am also grateful to Ra\'ul Carballo-Rubio, Stefano Liberati and Matt Visser for numerous discussions that inspired this work. Finally, I am indebted to Luca Buoninfante, Ra\'ul Carballo-Rubio, Carlos Barcel\'o, and Stefano Liberati for providing very useful feedback on this draft on a very short notice. 

\bibliographystyle{utphys}
\bibliography{refs}
\end{document}